\documentclass[showpacs,preprintnumbers,amsmath,amssymb]{revtex4}

\usepackage{graphicx}
\usepackage{dcolumn}
\usepackage{bm}

\begin{document}

\title{Nuclear collective excitations in Landau Fermi liquid theory}

\author{Bao-Xi Sun}

\affiliation{Institute of Theoretical Physics, College of Applied
Sciences, Beijing University of Technology, Beijing 100124, China}

\begin{abstract}
The nuclear collective excitations are studied within Landau Fermi
liquid theory. By using the nucleon-nucleon interaction of the
linear $\sigma-\omega$ model, the nuclear collective excitation
energies of different values of $l$ are obtained, which are fitted
with the centroid energies of the giant resonances of spherical
nuclei, respectively. In addition, it is pointed out that the
isovector giant resonances except $l=1$ correspond to the modes that
protons are in the creation state and neutrons are in the
annihilation state, and vice versa. Some mixtures of the nuclear
collective excitation states with different values of $l$ are
predicted.
\end{abstract}
\pacs{ 21.10.Re, 
21.30.Fe, 
21.65.-f, 
71.10.Ay  
  }
\maketitle

\section{Introduction}

The nuclear collective excitation states have been studied in the
framework of nuclear macroscopic and microscopic models. More
detailed descriptions on these models and references can be found in
Ref.~\cite{Greiner}. Nowadays, it is still a interesting topic in
nuclear physics. With several typical methods, such as the random
phase approximation with Skyrme interactions\cite{Zhang}, the
relativistic random phase approximation\cite{Cao,Ma,Ring09}, the
centroid energies and strength distributions of the giant resonances
of the nuclei are calculated and compared with the experimental
data.

Landau Fermi liquid theory is one of the important cornerstones of
traditional many-body theory in the condensed state physics. it is
very useful because it describes almost all known metals and many
non-metallic states, such as superconductors, anti-ferromagnetic
states, etc. However, this theory has not been used to solve nuclear
many-body problems successfully. In this work, I will try to
calculate  the collective excitation energies of the nuclear matter
within the framework of Landau Fermi liquid theory, and then compare
my calculation results with the experimental data of the nuclear
giant resonances of the nuclei.

This article is organized as follows:  in
Section~\ref{sect:Boltzmann}, the formalism on the reduced Boltzmann
equation is extended to the 3-dimensional of the Fermi liquid with
the spin taken into account. In Section~\ref{sect:Results}, The
calculation results on the nuclear collective excitation energies
are compared with the experimental data of the giant resonances of
nuclei. The conclusion is summarized in Section~\ref{sect:Summary}.

\section{Reduced Boltzmann equation of a Fermi liquid at zero temperature}
\label{sect:Boltzmann}

According to Ref.~\cite{Wen}, the Boltzmann equation with an
infinite quasi-particle lifetime for the nuclear matter can be
written as
\begin{equation}
\frac{\partial n_{\vec{k}\alpha}}{\partial t}+\frac{\partial
n_{\vec{k} \alpha}}{\partial \vec{x}} \cdot \frac{\partial
\tilde{\epsilon}}{\partial \vec{k}} - \frac{\partial n_{\vec{k}
\alpha}}{\partial \vec{k}} \cdot \frac{\partial
\tilde{\epsilon}}{\partial \vec{x}}=0,
\end{equation}
where $n_{\vec{k}\alpha}(\vec{x},t)={n_0}_{\vec{k}\alpha}+\delta
n_{\vec{k}\alpha}(\vec{x},t)$ is the occupation number of
quasi-nucleons, and $\tilde{\epsilon}$ denotes the quasi-nucleon
energy on the background of a collective excited state.

The quasi-nucleon density $ \tilde{\rho}_\alpha(\theta,\phi)$ and
the occupation $n_{\vec{k}\alpha}$ are related as follows:
\begin{equation}
\label{eq:qpden}
 \tilde{\rho}_\alpha(\theta,\phi)~=~\int \frac{k^2 dk}{(2\pi)^3}
 \delta n_{\vec{k} \alpha}
\end{equation}
with
\begin{equation}
\delta n_{\vec{k} \alpha}~=~\left\{
\begin{array}{c}
1,  \\
0.
\end{array}
\right.
\end{equation}
and $\alpha$ denotes the spin index. The isospin index is suppressed
in Eq.~(\ref{eq:qpden}).

In the momentum space, the linearized liquid equation of motion
takes the form
\begin{eqnarray}
\label{eq:liquid-theta-phi} i\frac{\partial}{\partial t}
\tilde{\rho}_\alpha(\theta,\phi,\vec{q},t)~=~q \sum_\beta \int
d\Omega^{\prime} \int d \Omega^{\prime
\prime}K(\theta,\phi;\theta^\prime,\phi^\prime)
M(\theta^\prime,\phi^\prime,\alpha;\theta^{\prime\prime},\phi^{\prime\prime},\beta)
\tilde{\rho}_\beta(\theta^{\prime\prime},\phi^{\prime
\prime},\vec{q},t),
\end{eqnarray}
where
\begin{equation}
K(\theta,\phi;\theta^\prime,\phi^\prime)~=~[\sin \theta \sin
\theta_q \cos(\phi-\phi_q)~+~\cos \theta \cos \theta_q]\frac{1}{\sin
\theta^\prime }\delta(\theta-\theta^\prime)\delta(\phi-\phi^\prime)
\end{equation}
with $(\theta_q, \phi_q)$ the angle of the momentum $\vec{q}$, and
\begin{equation}
M(\theta,\phi,\alpha;\theta^\prime,\phi^\prime,\beta)~=~v^\ast_F
\frac{1}{\sin \theta^\prime} \delta_{\alpha \beta}
\delta(\theta-\theta^\prime)\delta(\phi-\phi^\prime)
~+~\frac{k^2_F}{(2\pi)^3}f(k_F, \theta,\phi,\alpha; k_F,
\theta^\prime,\phi^\prime,\beta) .
\end{equation}
with$f(k_F,\theta,\phi,\alpha; k_F,
\theta^\prime,\phi^\prime,\beta)$ the Fermi liquid function, $k_F$
the Fermi momentum and $v^\ast_F$ the Fermi velocity.

In order to determine the form of Fermi liquid function, we study
the effective interaction between two nucleons. According to the
linear $\sigma$-$\omega$ model, the nucleons $\psi$ interact with
scalar mesons $\sigma$ through a Yukawa coupling $\bar
\psi\psi\sigma$ and with neutral vector mesons $\omega$ that couple
to the conserved baryon current $\bar \psi \gamma_\mu
\psi$\cite{Walecka}. the Lagrangian density can be written as
\begin{eqnarray}
\label{eq:Lagr}
 {\cal L}~&=&~\bar\psi
\left(i\gamma_{\mu}\partial^{\mu} -
M_N\right)\psi~+~\frac{1}{2}\partial_\mu\sigma\partial^\mu\sigma-\frac{1}{2}
m^2_\sigma \sigma^2_{} -\frac{1}{4}\omega_{\mu\nu}\omega^{\mu\nu}+
\frac{1}{2}
m^2_\omega\omega_\mu\omega^\mu \nonumber \\
&&-g_\sigma\bar\psi\sigma\psi-g_\omega\bar\psi \gamma_\mu \omega^\mu
\psi,
\end{eqnarray}
with $M_N$,$m_\sigma$ and $m_\omega$ the nucleon, scalar meson and
vector meson masses, respectively, and
$\omega_{\mu\nu}~=~\partial_\mu\omega_\nu-\partial_\nu\omega_\mu$
the vector meson field tensor.

\begin{figure*}
\includegraphics{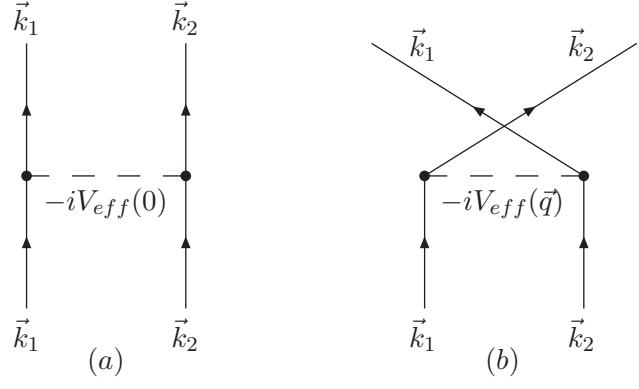}
\caption{\label{fig:fermi-liquid-f}The Fermi liquid function.$(a)$
Direct interaction; $(b)$ Exchange interaction with
$\vec{q}=\vec{k}_1-\vec{k}_2$.}
\end{figure*}

The effective nucleon-nucleon potential in the static limit can be
deduced directly with the Lagrangian in Eq.~(\ref{eq:Lagr})
\begin{equation}
\label{eq:veff} V_{eff}(\vec{q})-
V_{eff}(\vec{q}^{~\prime})\delta_{\alpha \beta}
\end{equation}
with
\begin{equation}
 V_{eff}(\vec{q})~=~\frac{-g^2_\sigma}
{\vec{q}^{~2}+m^2_\sigma}~+~\frac{g^2_\omega}{\vec{q}^{~2}+m^2_\omega},
\nonumber
\end{equation}
where the first term $V_{eff}(\vec{q})$ in
Eq.~(\ref{eq:veff})denotes the direct interaction between nucleons
and the second term $V_{eff}(\vec{q}^{~\prime})\delta_{\alpha
\beta}$ the exchange interaction with $\alpha$ and $\beta$ the spins
of interacting nucleons. The Fermi liquid function is instantaneous
interaction potential between two nucleons near the Fermi surface
with momenta $\vec{k_1}$ and $\vec{k_2}$ scattering into two
nucleons with same momenta $\vec{k_1}$ and $\vec{k_2}$, which is
depicted in Fig.~\ref{fig:fermi-liquid-f}. By using
Eq.~(\ref{eq:veff}), the Fermi liquid function takes the form
\begin{eqnarray}
\label{eq:fermifun} f(\vec{k}_1, \alpha; \vec{k}_2,
\beta)&=&V_{eff}(0)-V_{eff}(\vec{k}_1-\vec{k}_2)~\delta_{\alpha
\beta } \nonumber  \\
&=&\left(\frac{-g^2_\sigma}
{m^2_\sigma}~+~\frac{g^2_\omega}{m^2_\omega}\right)-
\left(\frac{-g^2_\sigma}
{(\vec{k_1}-\vec{k_2})^{2}+m^2_\sigma}~+~\frac{g^2_\omega}{(\vec{k_1}-\vec{k_2})^{2}+m^2_\omega}\right)
\delta_{\alpha \beta}
\end{eqnarray}
with
\begin{equation}
\vec{k}_1~=~(k_F, \theta, \phi),~~~~\vec{k}_2~=~(k_F, \theta^\prime,
\phi^\prime),
\end{equation}
and
\begin{eqnarray}
(\vec{k_1}-\vec{k_2})^{2}&=&2 k^2_F \left\{
1~-~\left[\cos{\theta}\cos{\theta^\prime}~+~\sin{\theta}\sin{\theta^\prime}\cos{(\phi-\phi^\prime)}\right]\right\}
\nonumber \\
&=&2 k^2_F \left( 1~-~ \hat{\vec{k}}_1 \cdot \hat{\vec{k}}_2
\right).
\end{eqnarray}

The direct interaction potential in the Fermi liquid function in
Eq.~(\ref{eq:fermifun}) is constant and only contribute a ground
state energy correction of the nuclear matter. In the framework of
the relativistic mean-field approximation or relativistic Hartree
approximation, the nucleon Fermi energy can be written as
\begin{equation}
\varepsilon^\ast_F \simeq M_N +
\frac{k_F^2}{2M^\ast_N}+(-\frac{g_\sigma^2}{m_\sigma^2}+\frac{g_\omega^2}{m_\omega^2})
\sum_\gamma \frac{k_F^3}{6\pi^2},
\end{equation}
where $\gamma$ denotes the summation over spins and isospins of the
nucleon, then the nucleon Fermi velocity $v^\ast_F$ is obtained as
\begin{equation}
\label{eq:velocity} v^\ast_F= \frac{\partial
\varepsilon^\ast_F}{\partial
k_F}=\frac{k_F}{M^\ast_N}+(-\frac{g_\sigma^2}{m_\sigma^2}+\frac{g_\omega^2}{m_\omega^2})
\frac{k_F^2}{2\pi^2},
\end{equation}
where the second term is just the contribution from the direct
interaction $V_{eff}(0)$ of the Fermi liquid function in
Eq.~(\ref{eq:fermifun}). Thus in the following calculation, only the
exchange term is reserved in the Fermi liquid function.

The quasi-nucleon density in Eq.~(\ref{eq:liquid-theta-phi}) can be
expanded in spherical harmonics with time-dependent shape parameters
as coefficients:
\begin{equation}
\label{eq:rho_l}
\tilde{\rho}_\alpha(\theta,\phi,\vec{q},t)~=~\sum_{l,m}\tilde{\rho}_\alpha(l,m,\vec{q},t)Y^\ast_{l,m}(\theta,\phi).
\end{equation}
Similarly, the functions $K(\theta,\phi;\theta^\prime,\phi^\prime)$
and
$M(\theta^\prime,\phi^\prime,\alpha;\theta^{\prime\prime},\phi^{\prime\prime},\beta)$
can also be expanded as
\begin{equation}
K(\theta,\phi;\theta^\prime,\phi^\prime)~=~\sum_{l,m,l^\prime,m^\prime}
K(l,m;l^\prime,m^\prime)Y^\ast_{l,m}(\theta,\phi)~Y_{l^\prime,m^\prime}(\theta^\prime,\phi^\prime),
\end{equation}
and
\begin{equation}
M(\theta^\prime,\phi^\prime,\alpha;\theta^{\prime\prime},\phi^{\prime\prime},\beta)
~=~\sum_{l_1,m_1,l_2,m_2} M(l_1,m_1,\alpha; l_2, m_2,
\beta)Y^\ast_{l_1,m_1}(\theta^\prime,\phi^\prime)~Y_{l_2,m_2}(\theta^{\prime\prime},\phi^{\prime\prime}),
\end{equation}
respectively.

 Therefore, the liquid equation of motion in the basis of
spherical harmonics can be rewritten as
\begin{eqnarray}
\label{eq:liquid-theta-phi2}    i\frac{\partial}{\partial t}
\tilde{\rho}_\alpha(l,m,\vec{q},t) ~=~ q \sum_\beta
\sum_{l^\prime,m^\prime}
\sum_{l^{\prime\prime},m^{\prime\prime}}K(l,m;l^\prime,m^\prime)
M(l^\prime,m^\prime,\alpha; l^{\prime\prime}, m^{\prime\prime},
\beta)
\tilde{\rho}_\beta(l^{\prime\prime},m^{\prime\prime},\vec{q},t).
\end{eqnarray}

Because the energy spectrum does not depend on the direction of
$\vec{q}$, we can choose $\vec{q}$ to be in the direction of
 $\theta_q~=~0$ and $\phi_q~=~0$, and then the function
 $K(\theta,\phi;\theta^\prime,\phi^\prime)$ becomes
\begin{equation}
K(\theta,\phi;\theta^\prime,\phi^\prime)~=~\frac{\cos \theta}{\sin
\theta^\prime }\delta(\theta-\theta^\prime)\delta(\phi-\phi^\prime).
\end{equation}
In the spherical harmonics,
\begin{eqnarray}
K(l,m;l^\prime,m^\prime)&=&\int \frac{\cos \theta}{\sin
\theta^\prime }\delta(\theta-\theta^\prime)\delta(\phi-\phi^\prime)
Y_{l,m}(\theta,\phi)~Y^\ast_{l^\prime,m^\prime}(\theta^\prime,\phi^\prime)
\sin \theta^\prime d \theta^\prime d \phi^\prime \sin \theta d
\theta d \phi \nonumber \\
&=& \left(a_{lm} \delta_{l+1,l^\prime} ~+~a_{l-1,m}
\delta_{l-1,l^\prime} \right)\delta_{m,m^\prime}
\end{eqnarray}
with
\begin{equation}
a_{lm}~=~\sqrt{\frac{(l+1)^2-m^2}{(2l+1)(2l+3)}}, \nonumber
\end{equation}
and
\begin{eqnarray}
 M(l_1,m_1,\alpha; l_2, m_2,
\beta)&=&\int [v^\ast_F \frac{1}{\sin \theta^\prime} \delta_{\alpha
\beta} \delta(\theta-\theta^\prime)\delta(\phi-\phi^\prime)
~+~\frac{k^2_F}{(2\pi)^3}f(k_F,\theta,\phi,\alpha; k_F,
\theta^\prime,\phi^\prime,\beta)]  \nonumber \\
&&Y_{l_1,m_1}(\theta,\phi)~Y^\ast_{l_2,m_2}(\theta^{\prime},\phi^{\prime})
d \Omega d \Omega^{\prime} \nonumber \\
&=&v^\ast_F  \delta_{\alpha \beta} \delta_{l_1,l_2}
\delta_{m_1,m_2}~-~\frac{k^2_F}{(2\pi)^3}
f_F(l_1,m_1;l_2,m_2)\delta_{\alpha,\beta},
\nonumber \\
\end{eqnarray}
where the Fock term
\begin{eqnarray}
\label{eq:f_F} f_F(l_1,m_1;l_2,m_2)&=&\int
V_{eff}(\vec{k}_1-\vec{k}_2)
Y_{l_1,m_1}(\theta,\phi)~Y^\ast_{l_2,m_2}(\theta^{\prime},\phi^{\prime})
\sin \theta d \theta d \phi \sin \theta^\prime d \theta^\prime d
\phi^\prime \nonumber \\
&=&\int \left(\frac{-g^2_\sigma}
{(\vec{k_1}-\vec{k_2})^{2}+m^2_\sigma}~+~\frac{g^2_\omega}{(\vec{k_1}-\vec{k_2})^{2}+m^2_\omega}\right)
Y_{l_1,m_1}(\theta,\phi)~Y^\ast_{l_2,m_2}(\theta^{\prime},\phi^{\prime})
\nonumber \\ && \sin \theta d \theta d \phi \sin \theta^\prime d
\theta^\prime d
\phi^\prime \nonumber \\
&=&f_F(l_1,l_2) \delta_{l_1,l_2} \delta_{m_1,m_2}
\end{eqnarray}
would give a contribution to the nuclear collective excitation when
$l_1=l_2$ and $m_1=m_2$. Therefore, the liquid equation of motion of
the quasi-nucleon can be rewritten as
\begin{eqnarray}
\label{eq:liquid-2}    i\frac{\partial}{\partial t}
\tilde{\rho}_\alpha(l,m,\vec{q},t) &=&q \sum_{l^\prime} \left(a_{lm}
\delta_{l+1,l^\prime} ~+~a_{l-1,m} \delta_{l-1,l^\prime} \right)
\left(v^\ast_F-\frac{k^2_F}{(2\pi)^3}f_F(l^{\prime},l^{\prime
})\right)
\tilde{\rho}_\alpha(l^{\prime},m,\vec{q},t), \nonumber \\
\end{eqnarray}
or the matrix equation form
\begin{eqnarray}
\label{eq:matrixeq}    i\frac{\partial}{\partial t}
\tilde{\rho}_\alpha(l,m,\vec{q},t) &=&q \tilde{K} \tilde{M}
\tilde{\rho}_\alpha(l,m,\vec{q},t), \nonumber \\
\end{eqnarray}
with
 \begin{equation}
 \tilde{K}_{l,l^\prime}=\left(a_{lm} \delta_{l+1,l^\prime}
~+~a_{l-1,m} \delta_{l-1,l^\prime} \right)
\end{equation}
and
 \begin{equation}
\tilde{M}_{l^\prime,l}=\left(v^\ast_F-\frac{k^2_F}{(2\pi)^3}f_F(l^{\prime},l^{\prime
})\right)\delta_{l^\prime,l}.
\end{equation}
The stability of the Fermi liquid requires the diagonal matrix
elements of $\tilde{M}$ to be positive definite. Hence, all the
value of $f_F(l^{\prime},l^{\prime })$ must be less than $(2\pi)^3
v^\ast_F /k_F^2$, and we can write $\tilde{M}$ as $\tilde{M}=W W^T$.
Letting $u_\alpha=W^T \tilde{\rho}_\alpha$, then
Eq.~(\ref{eq:matrixeq}) becomes

\begin{eqnarray}
\label{eq:Schrodinger}    i\frac{\partial}{\partial t}
u_\alpha(l,m,\vec{q},t) &=&q W^T \tilde{K} W
u_\alpha(l,m,\vec{q},t)~=~Hu_\alpha(l,m,\vec{q},t), \nonumber \\
\end{eqnarray}
where the Hamiltonian
\begin{eqnarray}
\label{eq:hamilton} H_{l,l^\prime}(m) &=&q(W^T \tilde{K}
W)_{l,l^\prime}~=~ q \left(a_{lm} \delta_{l+1,l^\prime} ~+~a_{l-1,m}
\delta_{l-1,l^\prime} \right)
\left(v^\ast_F-\frac{k^2_F}{(2\pi)^3}f_F(l,l)\right)^{1/2}\left(v^\ast_F-\frac{k^2_F}{(2\pi)^3}f_F(l^{\prime},l^{\prime
})\right)^{1/2}
\end{eqnarray}
is hermite and $H=H^\dagger$. The eigenvalues of $H$ would give us
the frequencies of the collective excitation modes of the nuclear
matter.

\section{Results}
\label{sect:Results}

In this section, the eigenvalues of the Hamiltonian in
Eq.~(\ref{eq:hamilton}) for different values of $l$ are calculated
with the Fermi liquid function in the linear $\sigma$-$\omega$
model. In Eq.~(\ref{eq:hamilton}),  Quantum number $m$ is fixed to
zero since our calculation will begin from $l=0$. The parameters in
Ref.~\cite{HS} are used in the calculation, i.e., $g_\sigma=10.47$,
$g_\omega=13.80$, $m_\sigma=520MeV$, $m_\omega=783MeV$ and
$M_N=939MeV$. Since the nucleon near the Fermi surface would be more
possible to be excited, we set the value of  nucleon momentum
$|\vec{q}|=k_F=1.36fm^{-1}$ in the calculation. When $f_F(l,l)=0$,
the Hamiltonian $H$ has a continuous spectrum and it generates the
particle-hole continuum of the nuclear matter in the relativistic
mean-field approximation. However, if the value of $f_F(l,l)$ is
large enough, in addition to the continuum eigenvalues, the spectrum
of $H$ has isolated positive and negative eigenvalues , and the
positive isolated eigenvalue corresponds to the energy of the
collective excitation of the nuclear matter with fixed $l$. However,
the negative eigenvalue of $H$ does not correspond to the negative
energy of the nuclear collective excitation modes. Actually, the
mode with a positive eigenvalue corresponds to the creation of a
nuclear collective excitation mode, while the the mode with a
negative eigenvalue corresponds to the annihilation of a nuclear
collective excitation mode.

\begin{figure*}
\includegraphics{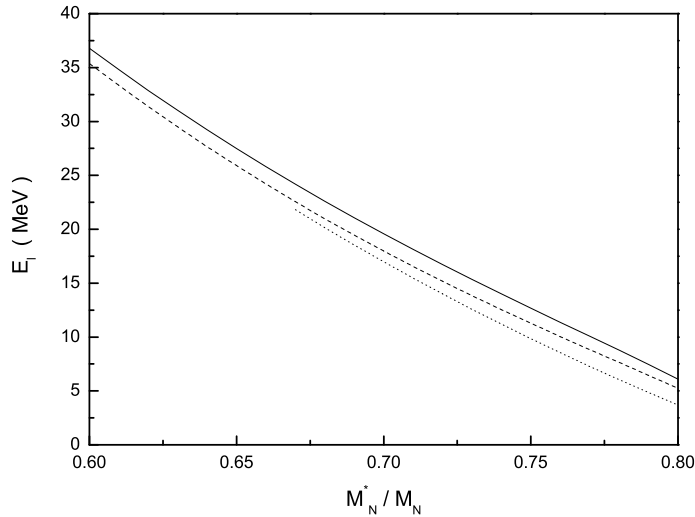}
\caption{\label{fig:e-m}The collective excitation energy $E_l$ for
the different value of $l$ as functions of the effective nucleon
$M^\ast_N$. l=0(Dash line), l=1(Solid line), l=2(Dot line)}
\end{figure*}

Fig.~\ref{fig:e-m} shows the collective excitation energy $E_l$ in
the nuclear matter versus the effective nucleon $M^\ast_N$. In my
calculation, the nuclear collective excitation energy is relevant to
the effective nucleon mass intensely, It can be seen that the
collective excitation energy decreases with the effective nucleon
mass increasing, and for $l=2$, when the effective nucleon mass is
less than $0.66M_N$, the isolated eigenvalues of $H$ can not be
generated.

Since the energy of the isospin scalar giant quadrupole resonance of
the nucleus $^{208}Pb$ is about $10.9\pm0.1$MeV\cite{Cao84}, the
effective nucleon mass can be fixed to be $M^\ast_N=0.742M_N$, which
generates an excitation collective energy of $10.92$MeV for $l=2$.
With the effective nucleon energy $M^\ast_N=0.742M_N$, the
collective excitation energies of the nuclear matter for different
values of $l$ are listed in Table~\ref{tab:e-l}. It is apparent that
the calculation results are fitted with the experimental values,
respectively.

\begin{table}[hbt]
\begin{center}
\begin{tabular}{c|cc}
 $l$ & $E_l~(MeV)$ & $E_{exp}~(MeV)$ \\
\hline
$0$ & $12.28$ & $14.17\pm0.28$ \\
$1$ & $13.73$ & $13.5\pm0.2$ \\
$2$ & $10.92$ & $10.9\pm0.1$ \\
\hline
\end{tabular}
\caption{ The collective excitation energies of the nuclear matter
for $l=0,1,2$ with the effective nucleon mass $M^\ast_N=0.742M_N$.
The corresponding experimental values for the excitation energies of
$^{208}Pb$ are also listed as $E_{exp}$,where the experimental value
for $l=0$ is taken from Ref.~\cite{Cao74}, the experimental value
for $l=1$ from Ref.~\cite{Cao83}, and the experimental value for
$l=2$ from Ref.~\cite{Cao84}. }
\label{tab:e-l}
\end{center}
\end{table}

The collective excitation of the nuclear matter with $l\ge3$, is
difficult to calculate in the framework of Landau Fermi liquid
theory with a Fermi liquid function deduced from the linear
$\sigma-\omega$ model. For the collective excitation  of the nuclear
matter with $l=3$, the positive isolated energy eigenvalue is
$3.12MeV$ with $M^\ast_N=0.8M_N$, which can be treated as the
low-energy octupole resonance\cite{Ring-Schuck}. However, the
high-energy octupole resonance can not be generated with our
model\cite{Cao79}.

\begin{figure*}
\includegraphics{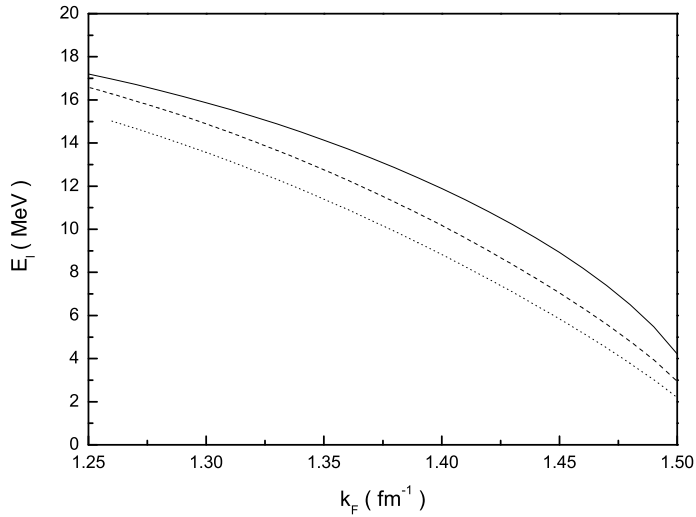}
\caption{\label{fig:e-kf}The collective excitation energy $E_l$ for
the different value of $l$ as functions of the Fermi momentum $k_F$
with $M^\ast_N=0.742M_N$. l=0(Dash line), l=1(Solid line), l=2(Dot
line)}
\end{figure*}

The nuclear collective excitation energy $E_l$ as functions of the
Fermi momentum $k_F$ for $l=0,1,2$ are illustrated in
Fig.~\ref{fig:e-kf}. When the value of the effective nucleon mass is
fixed, the collective excitation energy decreases with the Fermi
momentum increasing. For the case of $l=2$, the isolated energy
levels can not be excited from the continuum quasi-nucleon energy
levels when the Fermi momentum $k_F$ is less than $1.26fm^{-1}$.

The nuclear isoscalar giant resonances actually correspond to the
nuclear collective excitations with different values of $l$.
However, the nuclear isovector giant resonances correspond to the
nuclear collective excitation states that the collective excitation
of protons is creating with the energy $E_S(l)$, while the
collective excitation of neutrons is annihilating with the energy
$E_S(l)$, and vice versa. Hence, the energy of the nuclear isovector
giant resonance is about twice of the corresponding isoscalar giant
resonance in the nuclear matter, i.e.,
\begin{equation}
\label{twotime} E_V(l)=E_S(l)-(-E_S(l))=2E_S(l).
\end{equation}
The experimental data on the nuclear giant resonances in
Ref.~\cite{Cao74,Cao83,Cao84,Cao80,Cao81,Cao82} demonstrate that the
relation between the energy of nuclear isovector giant resonance and
that of nuclear isoscalar giant resonance in Eq.~(\ref{twotime}) is
correct approximately except for the case $l=1$, which will be
discussed in detail in \ref{sect:l1}.  In follows, I will study the
giant resonance energies with different values of $l$ in the
nucleus.

\subsection{Nuclear giant monopole resonances}

The nuclear giant monopole mode, $l=0$. The spherical harmonic
$Y_{00}(\theta,\phi)$ is constant, so that a non-vanishing value of
$\rho_\alpha(0,0,\vec{q},t)$ corresponds to a change of the Fermi
momentum according to Eq.~(\ref{eq:rho_l}). The associated
excitation is the so-called ${\sl breathing~mode}$ of the nucleus.
Supposed the proton and neutron densities can be calculated
approximately:
 \begin{equation}
\rho_p = \rho_0 \frac{Z}{A},~~~~\rho_n = \rho_0 \frac{N}{A}.
\end{equation}
Thus the calculated energies for isoscalar and isovector giant
monopole resonances of nuclei $^{208}Pb$, $^{144}Sm$, $^{116}Sn$,
$^{90}Zr$, $^{40}Ca$ and their corresponding experimental values are
listed in Table~\ref{tab:monopole}. Since the Fermi momentum of
protons $k_F(p)$ is different from that of neutrons, the collective
excitation energies of protons and neutrons, $E_0(p)$ and $E_0(n)$,
are different from each other. It shows the calculation results of
the proton excitation energy for heavy nuclei, such as $^{208}Pb$,
$^{144}Sm$ and $^{116}Sn$, are fitted with the corresponding
experimental centroid energy of the nuclear isoscalar monopole
resonance $E^S_{exp}$, while for those light nucleus, such as
$^{90}Zr$ and $^{40}Ca$, the calculation results are less than those
experimental values, respectively. Moreover, the sum of the
excitation energies of protons and neutrons $E_0(p)+E_0(n)$ should
be fitted with the nuclear isovector giant monopole energy. For
$^{208}Pb$, it is just in the range of the experimental values.
However, for light nuclei,  such as $^{90}Zr$ and $^{40}Ca$,
Similarly to the cases of nuclear isoscalar giant monopole, the
values of $E_0(p)+E_0(n)$ are less than the centroid energy of the
nuclear isovector giant monopole. Because the value of the effective
nucleon mass is determined on the collective excitation energy of
$^{208}Pb$ for $l=2$, it can be believed that with a little smaller
effective nucleon mass, the calculation results for the nuclei
$^{90}Zr$ and $^{40}Ca$ can fit with the experimental values very
well. Actually, with $M^\ast_N=0.717M_N$, we can obtain
$E_0(p)=E_0(n)=15.58MeV$ and $E_0(p)+E_0(n)=31.16MeV$ for $^{40}Ca$,
and $E_0(p)=17.57MeV$, $E_0(n)=13.13MeV$ and $E_0(p)+E_0(n)=30.7MeV$
for $^{90}Zr$, which are fitted with the corresponding experimental
centroid energies of the nuclear isoscalar and isovector giant
monopole resonances.

\begin{table}[hbt]
\begin{center}
\begin{tabular}{c|c|c|c|c|c|c|c}
 $l=0$ & $k_F(p)~(fm^{-1})$ & $k_F(n)~(fm^{-1})$ & $E_0(p)~(MeV)$ & $E_0(n)~(MeV)$
 & $E_0(p)+E_0(n)~(MeV)$ & $E^S_{exp}~(MeV)$ & $E^V_{exp}~(MeV)$\\
\hline
$^{208}Pb$ & $1.26$ & $1.45$ & $16.28$ & $7.05$ & $23.33$ & $14.17\pm0.28$ & $26.0\pm3.0$\\
$^{144}Sm$ & $1.29$ & $1.42$ & $15.26$ & $9.00$ & $24.26$ & $15.39\pm0.28$ & $-$ \\
$^{116}Sn$ & $1.29$ & $1.42$ & $15.26$ & $9.00$ & $24.26$ & $16.07\pm0.12$ & $-$ \\
$^{90}Zr$  & $1.31$ & $1.41$ & $14.50$ & $9.60$ & $24.10$ & $17.89\pm0.20$ & $28.5\pm2.6$ \\
$^{40}Ca$  & $1.36$ & $1.36$ & $12.28$ & $12.28$& $24.56$ & $-$             & $31.1\pm2.2$ \\
 \hline
\end{tabular}
\caption{The Fermi momenta and the $l=0$ collective excitation
energies of protons and neutrons for different nuclei with the
effective nucleon mass $M^\ast_N=0.742M_N$. The corresponding
experimental values for the nuclear isoscalar and isovector giant
monopole resonances  are labeled as $E^S_{exp}$ and $E^V_{exp}$,
where the experimental values for the nuclear isoscalar
 giant monopole resonances are taken from Ref.~\cite{Cao74}, the
experimental values for the nuclear isovector giant monopole
resonances from Ref.~\cite{Cao80,Cao81,Cao82}.}
\label{tab:monopole}
\end{center}
\end{table}

\subsection{Nuclear giant dipole resonances}
\label{sect:l1}

The dipole deformation of the nucleus is really a shift of the
center of mass. Thus the isospin isovector giant dipole resonance of
the nucleus actually corresponds to the creation of the $l=1$
collective excitation of protons or neutrons. The isoscalar giant
dipole resonance in $^{208}Pb$ with a centroid energy at
$E=22.5MeV$, using the $(\alpha,\alpha^\prime)$ cross sections at
forward angles\cite{Cao77}, should be a compression mode, which
corresponds to a creation of the $l=1$ collective excitation of
protons or neutrons and an annihilation of the $l=1$ collective
excitation of neutrons or protons simultaneously. The calculation
results and the corresponding experimental centroid energies are
listed in Table~\ref{tab:dipole}. For the heavy nucleus, $^{208}Pb$,
the calculation excitation energy for protons $E_1(p)$ with the
effective nucleon mass $M^\ast_N=0.742M_N$ is larger than the
experimental value, especially the sum $E_1(p)+E_1(n)$ is larger
than the corresponding energy of the giant isovector dipole
resonance of $^{208}Pb$. If we increase the value of the effective
nucleon mass to $M^\ast_N=0.755M_N$, we can obtain
$E_1(p)=15.53MeV$, $E_1(n)=6.57MeV$ and $E_1(p)+E_1(n)=22.1MeV$.
However, for the light nucleus, $^{40}Ca$, the excitation energies
of protons and neutrons are less than the experimental value, and we
must reduce the value of the effective nucleon mass to obtain a
correct excitation energy. With $M^\ast_N=0.70M_N$, we obtain
$E_1(p)=E_1(n)=19.58MeV$, and the sum $E_1(p)+E_1(n)=39.16MeV$ for
$^{40}Ca$. It is apparent that in order to obtain a more correct
excitation energy, the effective nucleon mass must take a larger
value for heavy nuclei, but a smaller value for light nuclei.

\begin{table}[hbt]
\begin{center}
\begin{tabular}{c|c|c|c|c|c|c|c}
 $l=1$ & $k_F(p)~(fm^{-1})$ & $k_F(n)~(fm^{-1})$ & $E_1(p)~(MeV)$ & $E_1(n)~(MeV)$
 & $E_1(p)+E_1(n)~(MeV)$ & $E^S_{exp}~(MeV)$ & $E^V_{exp}~(MeV)$\\
\hline
$^{208}Pb$ & $1.26$ & $1.45$ & $16.97$ & $8.93$ & $25.9$   & $22.5$ & $13.5\pm0.2$  \\
$^{90}Zr$  & $1.31$ & $1.41$ & $15.56$ & $11.37$ & $26.93$ & $-$    & $16.5\pm0.2$  \\
$^{40}Ca$  & $1.36$ & $1.36$ & $13.73$ & $13.73$& $27.46$  & $-$    & $19.8\pm0.5$  \\
 \hline
\end{tabular}
\caption{The Fermi momenta and the $l=1$ collective excitation
energies of protons and neutrons for different nuclei with the
effective nucleon mass $M^\ast_N=0.742M_N$. The corresponding
experimental values for the nuclear isoscalar and isovector giant
dipole resonances  are labeled as $E^S_{exp}$ and $E^V_{exp}$, where
the experimental value for the nuclear isoscalar
 giant dipole resonances are taken from Ref.~\cite{Cao77}, the
experimental value for the nuclear isovector giant dipole resonances
from Ref.~\cite{Cao83}.}
\label{tab:dipole}
\end{center}
\end{table}

\subsection{Nuclear giant quadrupole resonances}

The calculation results and the corresponding experimental centroid
energies of the giant quadrupole resonances of different nuclei are
listed in Table~\ref{tab:quadrupole}. The experimental value of the
isovector giant quadrupole resonance energy is just twice of the
isoscalar giant quadrupole resonance energy for $^{208}Pb$, it
manifests my prediction on the relation between the nuclear
isovector giant resonance and the corresponding isoscalar giant
resonance in Eq.~(\ref{twotime}) is correct. Actually, the
experimental values for the other nuclei, and for the monopole giant
resonance are also fitted with the relation in Eq.~(\ref{twotime}).
approximately. In Table~\ref{tab:quadrupole}, Since the average
neutron density is larger than the saturation density of the nuclear
matter, the collective excitation energies of the nuclei $^{208}Pb$
and $^{90}Zr$ for $l=2$ are smaller than 10MeV, they are
corresponding to the low-lying excitation states in the heavy
nuclei, which are in the range of $2-6$MeV for
$^{208}Pb$\cite{Ring-Schuck}. For light nuclei, such as $^{40}Ca$
and $^{16}O$, the collective excitation energies are less than the
corresponding experimental values. We can choice the effective
nucleon mass as $M^\ast_N=0.69M_N$, and obtain the values of
$E_2(p)=E_2(n)=18.54MeV$ and $E_2(p)+E_2(n)=37.08MeV$, which are
close to the experimental values of the isoscalar and isovector
giant quadrupole resonances of $^{40}Ca$, respectively.

\begin{table}[hbt]
\begin{center}
\begin{tabular}{c|c|c|c|c|c|c|c}
 $l=2$ & $k_F(p)~(fm^{-1})$ & $k_F(n)~(fm^{-1})$ & $E_2(p)~(MeV)$ & $E_2(n)~(MeV)$
 & $E_2(p)+E_2(n)~(MeV)$ & $E^S_{exp}~(MeV)$ & $E^V_{exp}~(MeV)$\\
\hline
$^{208}Pb$ & $1.26$ & $1.45$ & $15.02$ & $5.84$ & $20.86$  & $10.9\pm0.1$  & $22$\\
$^{90}Zr$  & $1.31$ & $1.41$ & $13.16$ & $8.27$& $21.43$ & $14.41\pm0.1$ & $-$ \\
$^{40}Ca$  & $1.36$ & $1.36$ & $10.92$ & $10.92$& $21.84$ & $17.8\pm0.3$  & $32.5\pm1.5$ \\
$^{16}O $  & $1.36$ & $1.36$ & $10.92$ & $10.92$& $21.84$ & $20.7$  & $-$ \\
 \hline
\end{tabular}
\caption{The Fermi momenta and the $l=2$ collective excitation
energies of protons and neutrons for different nuclei with the
effective nucleon mass $M^\ast_N=0.742M_N$. The corresponding
experimental values for the nuclear isoscalar and isovector giant
quadrupole resonances  are labeled as $E^S_{exp}$ and $E^V_{exp}$,
where the experimental value for the nuclear isoscalar
 giant quadrupole resonances are taken from Ref.~\cite{Cao84}, the
experimental value for the nuclear isovector giant quadrupole
resonances from Ref.~\cite{Cao78,Cao}.}
\label{tab:quadrupole}
\end{center}
\end{table}

\subsection{Mixture of different $l$ states}

From Eq.~(\ref{eq:Schrodinger}), it is apparent that there are four
kinds of independent collective excitations in the nuclei or nuclear
matter if the spin and isospin are both taken into account.
Therefore, protons and neutrons, even the protons or neutrons with
different spins can lie in different collective excitation states,
respectively. Here I will discuss another kind of mixture of
different $l$ states for protons or neutrons with the same spin. If
the exchange interactions $f_F(l,l)$ for different values of $l$ are
taken into account simultaneously, we will obtain the collective
excitation energies of the mixture of different $l$ states, which
are shown in Table~\ref{tab:mix}. It can be seen that the energy
values of the mixture are all greater than the excitation energies
with the determined $l$, which are components of the mixture state.
For the mixture of $l=0,1$, $l=0,1,2$ and $l=0,1,2,3$ states, the
mixture energies is about 20MeV, higher than the experimental
centroid energy values of the isoscalar giant resonances. Obviously
the contributions from the $l=0$ and $l=1$ states are important to
increase the mixture energy.

\begin{table}[hbt]
\begin{center}
\begin{tabular}{c|c}
 $l$ & $E~(MeV)$  \\
\hline
$0,1$ & $19.14$  \\
$0,2$ & $12.91$  \\
$1,2$ & $14.88$  \\
$0,1,2$ & $19.89$  \\
$0,1,2,3$ & $19.91$  \\
\hline
\end{tabular}
\caption{ The collective excitation energy of the mixture of
different $l$ states for the nuclear matter with the effective
nucleon mass $M^\ast_N=0.742M_N$ and the Fermi momentum
$k_F=1.36fm^{-1}$.}
\label{tab:mix}
\end{center}
\end{table}

\section{Summary}
\label{sect:Summary}

The method on Landau Fermi liquid theory in Ref.~\cite{Wen} is
extended to the 3-dimensional Fermion system with the spin
considered, and then by using the effective Lagrangian of the linear
$\sigma-\omega$ model, the Fermi liquid function is obtained  and
the nuclear collective excitation energies of different values of
$l$ are calculated within the framework of Landau Fermi liquid
theory. The results shows the nuclear collective excitation energies
decrease with the effective nucleon mass and the Fermi momentum
increasing. When the effective nucleon mass takes the value of
$0.742M_N$, and the Fermi momentum $k_f=1.36fm^{-1}$, the calculated
collective excitation energies of the nuclear matter for different
values of $l$ are fitted with the experimental values very well. In
addition, the isoscalar and isovector giant resonances of spherical
nuclei are studied within Landau Fermi liquid theory. we find the
centroid energies of the isoscalar giant resonances just correspond
to the positive isolated energy levels of the nuclear collective
excitation with different values of $l$, respectively, while the
isovector giant resonances except $l=1$ correspond to the modes that
protons(neutrons) are in the creation state of the collective
excitation and neutrons(protons) are in the annihilation state of
the same $l$. Furthermore, some mixtures of the collective
excitation states with different values of $l$, which have higher
energies, are predicted.

\section*{Acknowledgments}

This work is supported by the National Natural Science Foundation of
China under grant number 10775012.


\begin{thebibliography}{50}

\bibitem{Greiner}
W. Greiner and J. A. Maruhn, {\it Nuclear Models}, Springer-Verlag,
Berlin, 1996.

\bibitem{Zhang}
  I.~Hamamoto, H.~Sagawa and X.~Z.~Zhang,
  Phys.\ Rev.\  C {\bf 57}, R1064 (1998).

\bibitem{Cao}
L. G. Cao, Ph.D. Dissertation (2002).

\bibitem{Ma}
  Z.~Y.~Ma, A.~Wandelt, N.~Van Giai, D.~Vretenar, P.~Ring and L.~G.~Cao,
  Nucl.\ Phys.\  A {\bf 703}, 222 (2002).


\bibitem{Ring09}
  J.~Daoutidis and P.~Ring,
  Phys.\ Rev.\  C {\bf 80}, 024309 (2009).

\bibitem{Wen}
X. G. Wen, {\it Quantum Field Theory of Many-Body Systems}, Oxford
University Press, Oxford, 2004.

\bibitem{Walecka}
  J. D. Walecka, {\it Theoretical Nuclear and Subnuclear Physics}, 2nd Edition, World Scientific
  Press, Singapore, 2004.

\bibitem{HS} C. J. Horowitz and B. D. Serot, Nucl. Phys. A {\bf 368},
503 (1981).

\bibitem{Cao84} A. Van de Woude, Prog. Part. Nucl. Phys. {\bf 18},
217 (1987).

\bibitem{Cao74}
  D.~H.~Youngblood, H.~L.~Clark and Y.~W.~Lui,
  Phys.\ Rev.\ Lett.\  {\bf 82}, 691 (1999).

\bibitem{Cao83} B. L. Berman and S. C. Fultz, Rev. Mod. Phys. {\bf
47},  713 (1975).

\bibitem{Ring-Schuck} P. Ring and P. Schuck, {\it The Nuclear Many-Body
Problem}, Springer-Verlag, Berlin, 2004.

\bibitem{Cao79}
  T.~Yamagata {\it et al.},
  Phys.\ Rev.\  C {\bf 23}, 937 (1981)
  [Erratum-ibid.\  C {\bf 23}, 2798 (1981)].

\bibitem{Cao80}
  A.~Erell {\it et al.},
  Phys.\ Rev.\ Lett.\  {\bf 52}, 2134 (1984).

\bibitem{Cao81}
  A.~Erell {\it et al.},
  Phys.\ Rev.\  C {\bf 34}, 1822 (1986).

\bibitem{Cao82} J. D. Bowman, {\it Nuclear Structure},
North-Holland, Amsterdam, 1985. P.549.

\bibitem{Cao77}
  B.~F.~Davis {\it et al.},
  Phys.\ Rev.\ Lett.\  {\bf 79}, 609 (1997).

\bibitem{Cao78}
  D.~M.~Drake, K.~Aniol, I.~Halpern, S.~Joly, L.~Nilsson, D.~Storm and S.~A.~Wender,
  Phys.\ Rev.\ Lett.\  {\bf 47}, 1581 (1981).

\end{thebibliography}
\end{document}